# Hexagonal Be*X* (*X*: S, Te) monolayer as potential electrode material for alkali metal-ion batteries: A DFT perspective

Hetvi Jadav, Sadhana Matth, Himanshu Pandey*

*Condensed Matter & Low-Dimensional Systems Laboratory, Department of Physics, Sardar Vallabhbhai National Institute of Technology, Surat, Gujarat, 395007, India*

*Corresponding Author E-mail: hp@phy.svnit.ac.in

**Abstract**

Metal-ion batteries (MIBs) are essential for transitioning to a cleaner and more sustainable energy future. By employing the density functional formalism, we have investigated the hexagonal (*h*) monolayer of BeS and BeTe as electrode materials for alkali (Li and Na) MIBs. The structural and thermodynamic stability, adsorption of Li/Na atoms, density of states, diffusion and migration of atoms, and capacity are systematically investigated. The structure of *h*-BeS and *h*-BeTe remains stable upon adsorption of the adatoms, with improved electronic conductivity of these monolayers. The climbing image-nudged elastic band calculations estimate a low diffusion barrier of 0.16 eV (0.01 eV) for Li (Na) in *h*-BeS and 0.20 eV (0.16 eV) for Li (Na) in *h*-BeTe. Also, a maximum storage capacity of 580 mAh g$^{-1}$ for Li and 1305 mAh g$^{-1}$ for Na in *h*-BeS and 174 mAh g$^{-1}$ for *h*-BeTe for both metal ions is estimated.

1. **Introduction**

The increased consumption and limited resources of conventional fuels (natural gas, coal, and oil) have impacted the climate globally and harmed human lives and health. Even while fossil fuels can continue to provide for human requirements in the twenty-first century, it is essential to remember that resources will eventually run out [1-3]. Renewable energy may be successfully integrated into the system by using effective energy storage technology. Furthermore, effective power storage technologies are necessary for (all-electric or hybrid) vehicles, which could decrease associated pollutants and the usage of fossil fuels in transportation. Finally, the increasing need for portable consumer gadgets that are lighter and smaller necessitates the development of increasingly effective portable electricity storage solutions [4-5].

The Li-ion secondary batteries (LIBs) are the vital components of electric vehicles and portable electronics, among other energy devices [6-8]. However, due to limited sources of lithium, its high cost and safety concerns limit the future advancements of LIBs, especially in large-scale applications [4,5,9]. So, various other ions such as Na, K, Mg, and Ca are being examined due to the greater availability and comparable electrochemical principles to Li [9-17]. Out of these, the Na-ion batteries (NIBs) [10,12,18] are anticipated to be another option in place of LIBs in the future, as there are plenty of natural deposits of Na and a similar storage process. Apart from these, K-ion batteries (KIBs), another type of alkali ion-based battery, have also drawn much interest recently [11,19-21]. As is generally known, metal ions move between the anode and cathode through an electrolyte in all metal-ion batteries (MIBs). The issue is that not all MIBs can use the anode materials which is being used for LIBs [21,22]. To meet the requirements for a power system, for instance, the currently being used graphite anode, which offers a restricted storage capacity of 372 mA h g$^{-1}$ for LIBs [23], but provides a relatively low capacity of 150 mA h g$^{-1}$ and 273 mA h g$^{-1}$ for NIBs [24] and KIBs [25], respectively. Therefore, researchers have explored many materials that could be used in NIBs and KIBs. Among the various candidates for anode, many 2$D$ materials, such as BeB$_2$, BeC$_7$, Be$_2$C, and Be$_2$B$_2$, exhibit a high storage capacity in the range of 1350 to 2300 mA h g$^{-1}$ for LIBs [26-30], with no structural deterioration, which generally occurs during the adsorption and desorption of Li atoms. Specifically, the semiconducting Be$_2$B monolayer produces the largest capacity ever measured for magnesium ion batteries, 7436 mA h g$^{-1}$ [28].

In the present work, we theoretically studied the performance of $h$-BeS and $h$-BeTe as electrode materials for alkali MIBs. Applying density functional formalism, we have investigated structural, dynamical, mechanical and thermal stabilities, then studied the adsorption of Li/Na atoms to these monolayers, and finally, the diffusion barrier energy, open circuit voltage (*OCV*) and maximum storage capacity are estimated. Our results indicate that these monolayers possess low negative adsorption energy and diffusion barriers, which are suitable for use as electrodes for MIBs. Most importantly, we observed that the BeS monolayer has high storage capacities for both metal atoms, Li and Na.

## 2. Computational Details

In this work, an *ab initio* investigation is carried out on 3×3×1 monolayers of $h$-BeS and $h$-BeTe for MIBs, using the Quantum Espresso package [31] under PBE-GGA exchange–correlation functional [32]. The *cutoffs* for the kinetic energy and charge density of 50 Ry and 300 Ry for BeS were taken, whereas those of 50 Ry and 500 Ry were taken for BeTe. Broyden-Fletcher-Goldfarb-Shanno minimization technique was implemented in variable-cell relaxation calculations to optimize the crystal structure with the iterated self-consistent field (*scf*) method using an optimized *k*-point mesh of 4×4×1. In contrast, electronic properties are calculated for a dense grid of 16×16×1. Grimme's dispersion approach (DFT-D2) is considered for the dispersion corrections and to enhance the adsorption energy predictions. Along the *c*-axis, a 20 Å vacuum is taken to avoid inter-layer interactions. The metal-ion diffusion barrier energy is calculated using the climbing image nudged elastic band (CI-NEB) [33] approach to find the energy-minimum route between the specified beginning and final configurations. The *ab initio* molecular dynamics (AIMD) simulations were carried out using the Vienna Ab initio Simulation Package (VASP) [34]. The BoltzTraP2 [35] software package was used to calculate the electrical conductivity ($\sigma$) to relaxation time ($\tau$) ratio ($\sigma/\tau$). A denser k-point grid of $40 \times 40 \times 1$ was utilized to ensure accurate estimation of $\sigma/\tau$.

## 3. Results and Discussion
### 3.1 Structure optimization and its stability

A 3×3×1 supercell, which includes nine Be atoms and nine S/Te atoms, is considered for the calculations. The optimized monolayers of *h*-BeS and *h*-BeTe exhibit a honeycomb planar structure; the top and side views of them are shown in Fig. 1 (a). Figure 1 (b) shows the calculated ground state *scf* energy variation with the planar lattice parameter for BeS and BeTe monolayers. The equilibrium lattice parameters (bond length) are estimated to be 3.45 Å (1.99 Å) and 4.02 Å (2.32 Å), respectively, for *h*-BeS and *h*-BeTe unit cells with a bond angle of 120° between the Be and S/Te atoms. These values are comparable to those reported for Be-based hexagonal monolayers [36-39]. Table 1 summarizes the optimized structural parameters of these monolayers. Estimating formation energy ($E_f$) is a vital indicator to assess the thermodynamic stability. The $E_f$ values at 0 K can be obtained by applying $E_f = E_{h-BeS/Te} - E_{Be} - E_{S/Te}$, where $E_{h-BeS/Te}$, $E_{Be}$, and $E_{S/Te}$ are the energies of the *h*-BeS/Te monolayer, isolated Be atom, and isolated S/Te atom, respectively. Both monolayers are found to be thermodynamically stable and experimentally feasible in terms of synthesis, as indicated by negative values of $E_f$ for *h*-BeS and *h*-BeTe, which are around -9.67 eV/unit cell and -7.28 eV/unit cell, respectively.

**Table 1**: Optimized parameters for *h*-BeS and *h*-BeTe monolayers.

|  | *h*-BeS | | *h*-BeTe | |
| --- | --- | --- | --- | --- |
|  | **This work** | **Earlier work** | **This work** | **Earlier work** |
| **Lattice parameter (Å)** | 3.45 | 3.459 [36]<br>3.46 [37]<br>3.440 [38] | 4.02 | 4.03 [37] |
| **Bond Length (Å)** | 1.99 | 1.997 [36]<br>1.994 [39] | 2.32 | - |
| **Bandgap (eV)** | 4.61 | 4.26 [38]<br>4.469 [39] | 3.05 | 3.33 [37] |

By investigating the mechanical stability of a material, one can check its ability to maintain its internal cohesion and resistance against deformation or fracture under external stresses, such as force, pressure, and impact. For this, we have examined the elastic properties of pristine monolayers by using the standard algorithm as implemented in the *thermo_pw* code. Two independent second-order in-plane elastic constants, $C_{11}$ and $C_{12}$, are estimated, and another elastic constant ($C_{66}$) is deduced using the relation, $C_{66} = (C_{11}-C_{12})/2$. These elastic constants are tabulated in Table 2. BeS and BeTe are found to be mechanically stable as these elastic constants satisfy the Born–Huang stability criteria [40,41] for the hexagonal system, given by $C_{11} > 0$, $C_{11} - C_{12} > 0$, $C_{66} > 0$.

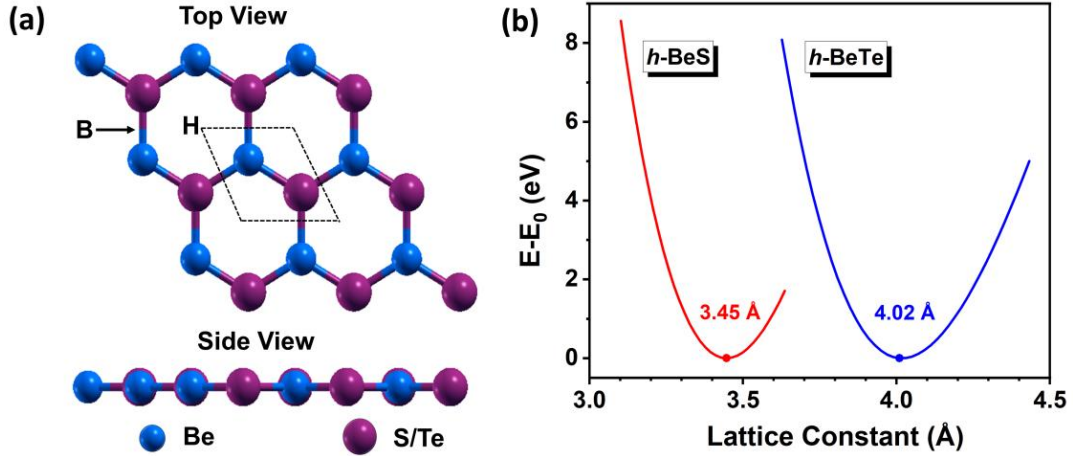

**Figure 1:** (a) Top and side view of a 3×3×1 supercell of *h*-BeS and *h*-BeTe; the black dashed parallelogram box represents the unit cell, and B represents the adsorption sites available at the bond between Be and S/Te atoms, whereas H indicates the hollow site. (b) Variation of ground state *scf* energy of *h*-BeS and *h*-BeTe monolayers with planar lattice parameter. Here, $E_0$ represents the corresponding energy minimum.

**Table 2:** Elastic constants $C_{11}$, $C_{12}$, and $C_{66}$ (in GPa) for *h*-BeS and *h*-BeTe monolayers.

| Monolayer | $C_{11}$ | $C_{12}$ | $C_{66}$ |
|---|---|---|---|
| BeS | 40.08 | 15.86 | 12.11 |
| BeTe | 29.07 | 10.67 | 9.20 |

Furthermore, to assess the dynamical stability of the monolayers considered in this study, phonon dispersion is calculated and plotted in Fig. 2. The absence of imaginary frequencies in the phonon dispersion of *h*-BeS and *h*-BeTe confirms the dynamic stability of both monolayers, ensuring the reliability of the optimized crystal structure. This structure, consisting of two atoms per unit cell, generates six distinct phonon branches. Three of these correspond to acoustic modes, which are typically the out-of-plane acoustic (ZA), transverse acoustic (TA), and longitudinal acoustic (LA) modes located in the low-frequency region. These LA and TA modes correspond to the vibrations of atoms in the longitudinal and transverse (within the lattice plane) directions of the wave propagation. In contrast, the ZA mode refers to the out-of-plane vibrations of atoms, meaning atoms move in a direction perpendicular to the plane of the monolayers. At the $\Gamma$ point, a degeneracy can be seen for these three acoustic modes, which is due to the symmetry of the lattice and because at $q = 0$, the phonon does not distinguish between different types of in-plane or out-of-plane vibrations due to the symmetry of the crystal structure. LA and TA modes typically split as the wavevector $q$ moves away from the $\Gamma$ point. The ZA mode remains much than the LA and TA modes. Apart from these acoustic modes, the remaining three branches are optical modes, located at higher frequencies and divided into two longitudinal optical modes (LO) and one transverse optical mode (TO). The non-degenerate TO mode occurs around 250 and 130 cm$^{-1}$, for BeS and BeTe monolayers, whereas the doubly degenerate LO modes occur around 750 and 600, respectively, for these monolayers.

In addition to the above, the temperature stability of these monolayers was examined via AIMD simulations [42], which were performed inside the NVT ensemble. The 3×3×1 supercell was used for these simulations, which were run at 300 K with a time step of 1 fs and a total simulation time scale of 5000 fs. Figure 3(a-f) displays the overall energy fluctuation as a function of time steps for pristine and metal-adsorbed monolayers, with the maximum adatoms, as these fully loaded layers are also considered while estimating *OCV*. As the total energy fluctuates around a stable average, we can assume that the simulations have attained equilibrium; hence, these multilayers have the potential to preserve their structure.

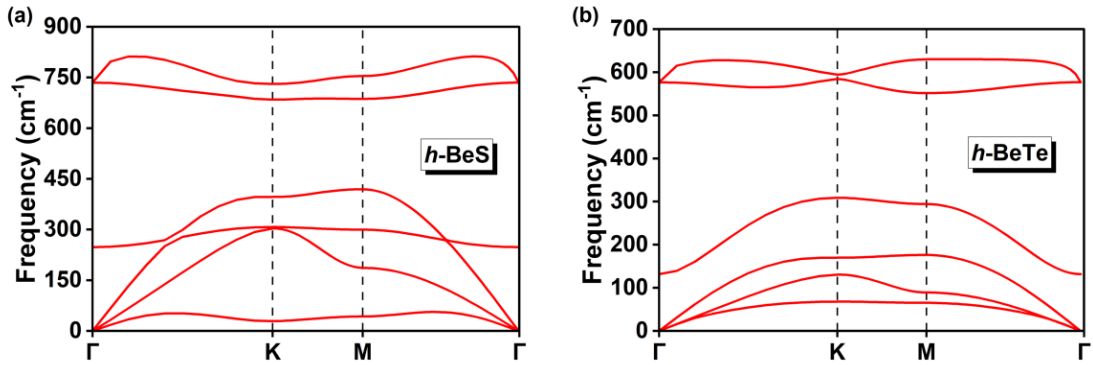

**Figure 2:** Phonon dispersion curves for (a) *h*-BeS and (b) *h*-BeTe nonolayers.

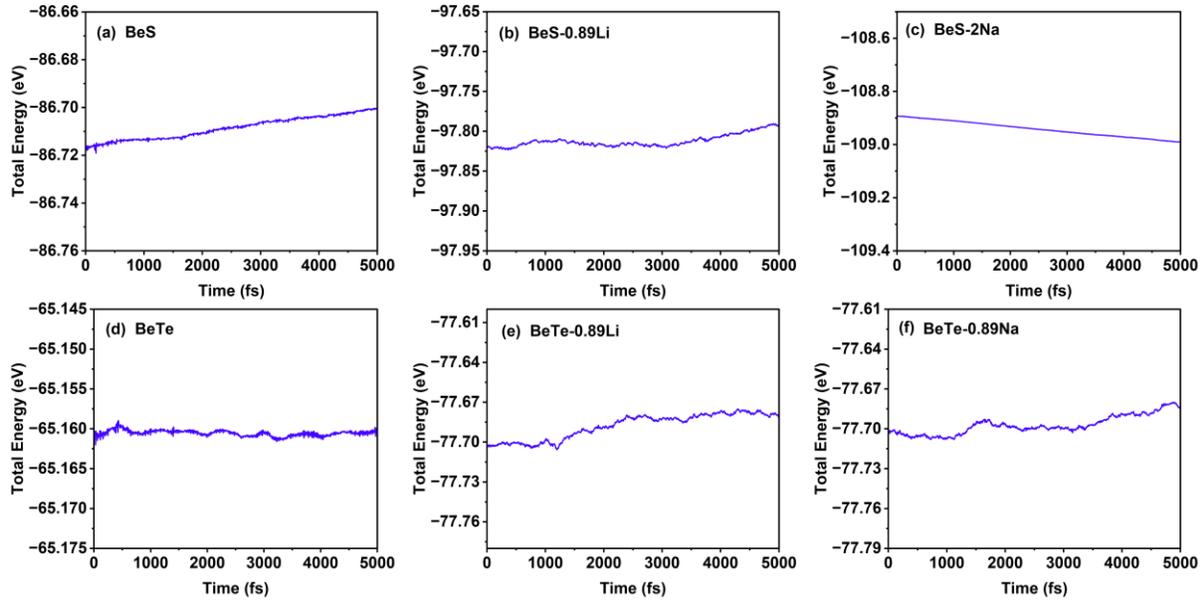

**Figure 3:** Total energy fluctuations during AIMD simulations at 300 K temperature for (a) BeS, (b) BeS-0.89Li, (c) BeS-2Na, (d) BeTe, (e) BeTe-0.89Li, and (f) BeTe-0.89Na systems.

## 3.2 Adsorption of alkali metal (Li/Na) ions on monolayers

Finding an appropriate adsorption site for the metal atom is crucial before further investigating any properties. Here, four typical adsorption sites are considered based on the geometry of BeS and BeTe monolayers. As marked in Fig. 1, the sites named as Be, B, H, and S/Te represent Be atom, Be–S/Te bond, hollow site, and S/Te atom, respectively. The ground state energies of the single metal adsorption systems are calculated with the DFT-D2 dispersion to understand the effect of van der Waals interactions. After the structural relaxation, it is found that the H-and S-sites have the lowest adsorption energy for the BeS monolayer adsorbed with Li and Na atoms, respectively. In contrast, the H-site is preferred for the adsorption of both the alkali metals for the BeTe monolayer. The adsorption energy ($E_{ad}$) is calculated using Eq. (1).

$$E_{ad} = (E_{BeS/Te-M} - E_{BeS/Te} - nE_M)/n \qquad (1)$$

Where $E_{BeS/Te-M}$, $E_{BeS/Te}$, and $E_M$ are the ground state energies of the Li/Na adsorbed monolayer, the respective pristine monolayer and isolated Li/Na atoms. And, $n$ is the number of adsorbed Li/Na atoms on the respective monolayer. Table 3 summarizes the $E_{ad}$ values calculated for the adsorption of a single alkali metal atom at different sites of BeS and BeTe monolayers. Negative values for $E_{ad}$ are found for all the cases, as the system lowers its total energy when the alkali metal binds to the monolayer, ensuring that the adsorption of Li/Na atoms is thermodynamically favorable rather than forming metal clusters [43,44]. Also, more negative values indicate a stronger interaction between the Li/Na atoms and the monolayers. From this, one can infer that for the adsorption of Li, the H-site is preferable for both monolayers, whereas for the adsorption of Na, S- and H-sites are found with the least $E_{ad}$ values for BeS and BeTe monolayers, respectively. As the alkali metal atoms have low ionization energies, they often donate electrons, suggesting ionic interaction or charge transfer to the monolayers. By looking at the magnitude of $E_{ad}$, the nature of the interaction can be inferred. Typically, for the physisorption (weak van der Waals forces) $E_{ad} \leq$ -0.5 eV, and for a weak chemisorption or ionic bonding, 0.5 $\leq E_{ad} \leq$ -2 eV. So, here, physisorption of Na atoms occurred while adsorbing on the BeS monolayer.

**Table 3:** Adsorption energy (in eV) for a single alkali metal atom at the hollow, bond, Be, and S/Te sites in both BeS and BeTe monolayers.

| Site | BeS | | BeTe | |
|---|---|---|---|---|
| | Li | Na | Li | Na |
| hollow (H) | -0.603 | -0.136 | -1.189 | -0.782 |
| bond (B) | -0.400 | -0.153 | -0.982 | -0.646 |
| Be | -0.538 | -0.144 | -1.099 | -0.601 |
| S/Te | -0.363 | -0.157 | -0.877 | -0.613 |

We also investigated how the interaction between the adatoms and monolayer changes with increasing the number of specific alkali atoms on different preferred adsorption sites for 3×3×1 supercell structure. Apart from a minute bending, no change in the lattice parameter is observed.

This bending is in the out-of-plane direction due to atoms shifting and occurs with the increased adsorption of alkali metal atoms. Hence, the optimized distance of the adsorbed adatoms with respect to the base monolayer is estimated, which is much larger than the average bending in the monolayer. The variation of average distance (in Å) with number of adatoms for Li/Na adsorbed BeS and BeTe monolayers is plotted in Fig. 4(a-d). Henceforth, any change in the volume due to progressive adsorption of alkali metal atoms can be understood in terms of the average optimized distance of adsorbed atoms. This variation is in the range of 0.1 to 2 Å at the respective adatom site (denoted by blue colour bar in Fig. 4), whereas the values of average optimized distance of adsorbed atoms for BeS and BeTe monolayers remain almost constant (with a fluctuation in the range 0.5 to 1 Å) with increasing the number of Li/Na adatoms. This suggests that the interaction between the adatoms and monolayer does not change with progressive adsorption, and the volume of the unit cell is almost constant, which is required for battery applications.

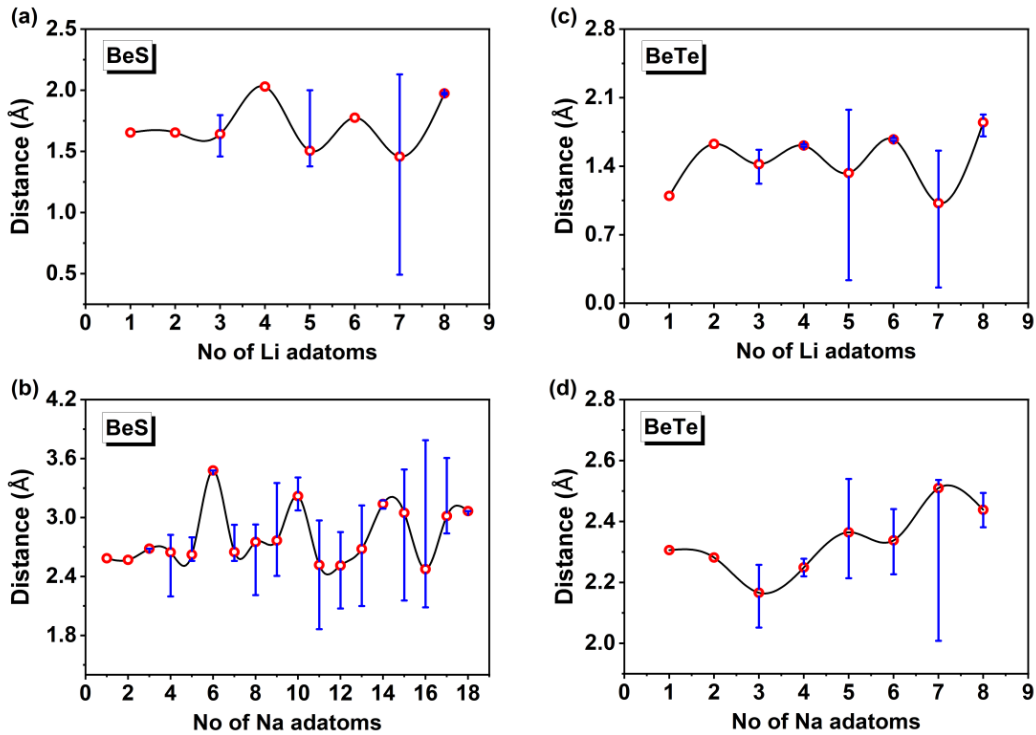

**Figure 4**: Average distance of adatoms from the monolayer (a) BeS-Li, (b) BeS-Na, (c) BeTe-Li and (d) BeTe-Na

### 3.3 Electronic properties

The better electronic conductivity of the electrode material is essential for any battery operation and has a longer lifespan. Hence, it is crucial to investigate the electronic structure and its correlation with battery performance. To comprehend the electronic character, band structure and corresponding density of state plots for pristine and metal-adsorbed monolayers are depicted in Figures 5 (a-c) and 5 (e-g), for BeS and BeTe monolayers, respectively. Here, zero on the $y$-axis represents the position of the Fermi level. These plots indicate the range of energy levels that an electron can occupy within the respective system and thus affect the electronic band gap and electrical conductivity. The pristine BeS and BeTe monolayers are found to be wide band gap semiconductors, [see Fig. 5 (a,d)] exhibiting indirect band gaps of 4.61 eV and 3.05 eV, respectively. These findings are consistent with earlier studies [37-39]. From these electronic structure calculations, we can infer that before adsorption, monolayers are semiconducting in nature, but after the adsorption of the alkali metal atoms, a metallic character is induced, as evident from the band crossing to the Fermi level. Here, the alkali metals easily donate their valence $s$-electron; this can be considered a doped monolayer, which shifts the Fermi level upward. With this, new states can be formed near the Fermi level, significantly influencing $\sigma$. To check the effect of adsorption of alkali metal atoms on the conductivity of the monolayers, electrical transport investigations were carried out using the BoltzTraP2 package. Figure 5 (d) and 4(h) show the temperature dependence of $\sigma/\tau$ for pristine and Li/Na-adsorbed monolayers. For pristine BeS and BeTe monolayers, a monotonous increase in $\sigma/\tau$ with temperature confirms the semiconducting behaviour. On the other hand, after the adsorption of Li and Na atoms to both BeS and BeTe monolayers, a general increase in the $\sigma/\tau$ values is found. This value also decreases with the temperature rise, ensuring a metallic nature after adsorption.

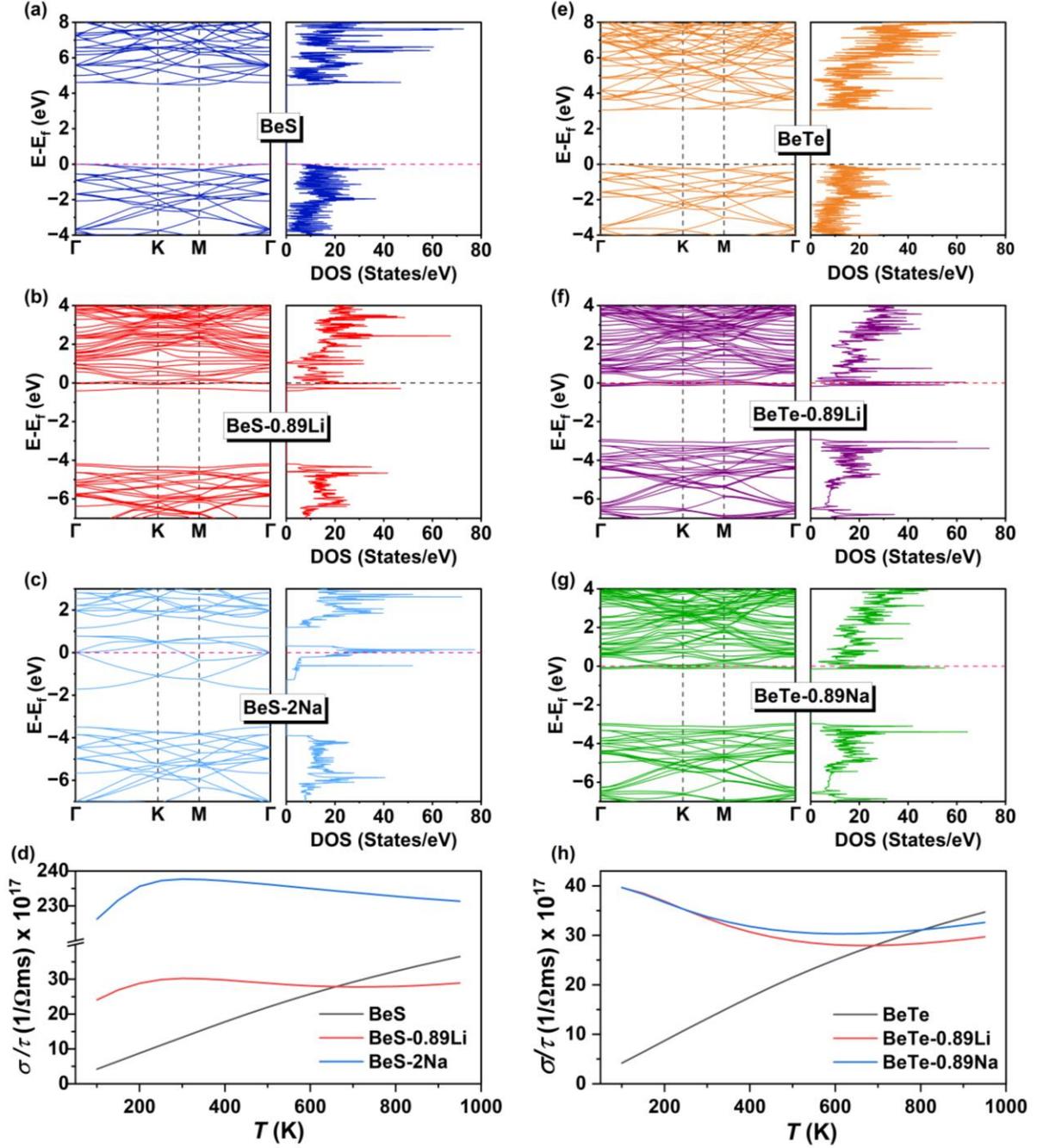

**Figure 5:** Band structure and density of state plots for different (a-c) BeS (e-g) BeTe-based systems. The electrical conductivity per relaxation times for corresponding (d) BeS and (h) BeTe-based systems.

### 3.4 Diffusion/migration of alkali metal atoms on monolayers

The mobility of alkali metal atoms across the electrodes significantly impacts the battery's charge-discharge rate, making it a key factor in evaluating their suitability as electrode materials.

Therefore, various possible diffusion pathways for Li and Na atoms on the adsorbed BeS and BeTe monolayers have been investigated and are illustrated in Fig. 6. To calculate the diffusion energy, CI-NEB calculations were performed, which is a reliable method for identifying the minimum energy path between defined initial and final states. Based on the optimized adsorption sites and structural configuration of both monolayers, several potential migration paths were considered, as shown in Fig. 6. For the migration of the Li atom in the BeS monolayer, three distinct diffusion routes connecting the adjacent hollow sites are taken, as depicted in Fig. 6 (a–c). The diffusion barriers for the paths H–S–H, H–H, and H–Be–H are 0.23 eV, 0.21 eV, and 0.16 eV, respectively. Therefore, H-S-H is the most preferable path among the three considered paths for migrating the Li atom on the BeS monolayer. Similarly, path S–Be–S is the potential diffusion channel for the Na atom on the BeS monolayer, as shown in Fig. 6 (d). One feasible migration path, H–Be–H, was found to migrate Li atoms on the BeTe monolayer [see Fig. 6 (e)]. However, three paths were again taken for the migration of the Na atom on the BeTe monolayer, which is depicted in Fig. 5(f-h). The diffusion energy barrier values for the paths H–Te–H, H–H, and H–Be–H are estimated to be 0.17 eV, 0.16 eV, and 0.18 eV, respectively. Thus, the path H–H, with low diffusion energy, is preferred for this scenario. Overall these calculated values of barrier energies are in the range 0.01–0.20 eV, which is lower than the previously reported values for Li and Na atom in aluminium carbide monolayer (for Li 0.78 eV and for Na 0.41 eV) [45]; $MoN_2$ (for Li 0.78 eV and for Na 0.56 eV) [46]; $P_3S$ (for Li 0.74 eV and for Na 0.26 eV) and $C_3S$ (for Li 0.68 eV and for Na 0.30 eV) [47]; $BeB_2$ (for Li 0.47 eV) [26]. The lower diffusion energy reported in this work can be advantageous for quick charge conduction and, hence, a viable option for electrode material for LIBs and NIBs.

### 3.5 *OCV* and Storage Capacity

The application of MIBs is greatly affected by their *OCV*, which depends on the choice of anode and cathode. So, it is essential to theoretically analyze the materials for anode and cathode for their use in MIBs before their realization in actual battery applications [26,43]. The charge-discharge mechanism in Li/Na adsorbed BeS and BeTe monolayers can be understood using the reactions:

$$BeS + xM^+ + xe^- \underset{Discharging}{\overset{Charging}{\rightleftarrows}} BeS + xM$$

$$BeTe + xM^+ + xe^- \underset{Discharging}{\overset{Charging}{\rightleftarrows}} BeTe + xM$$

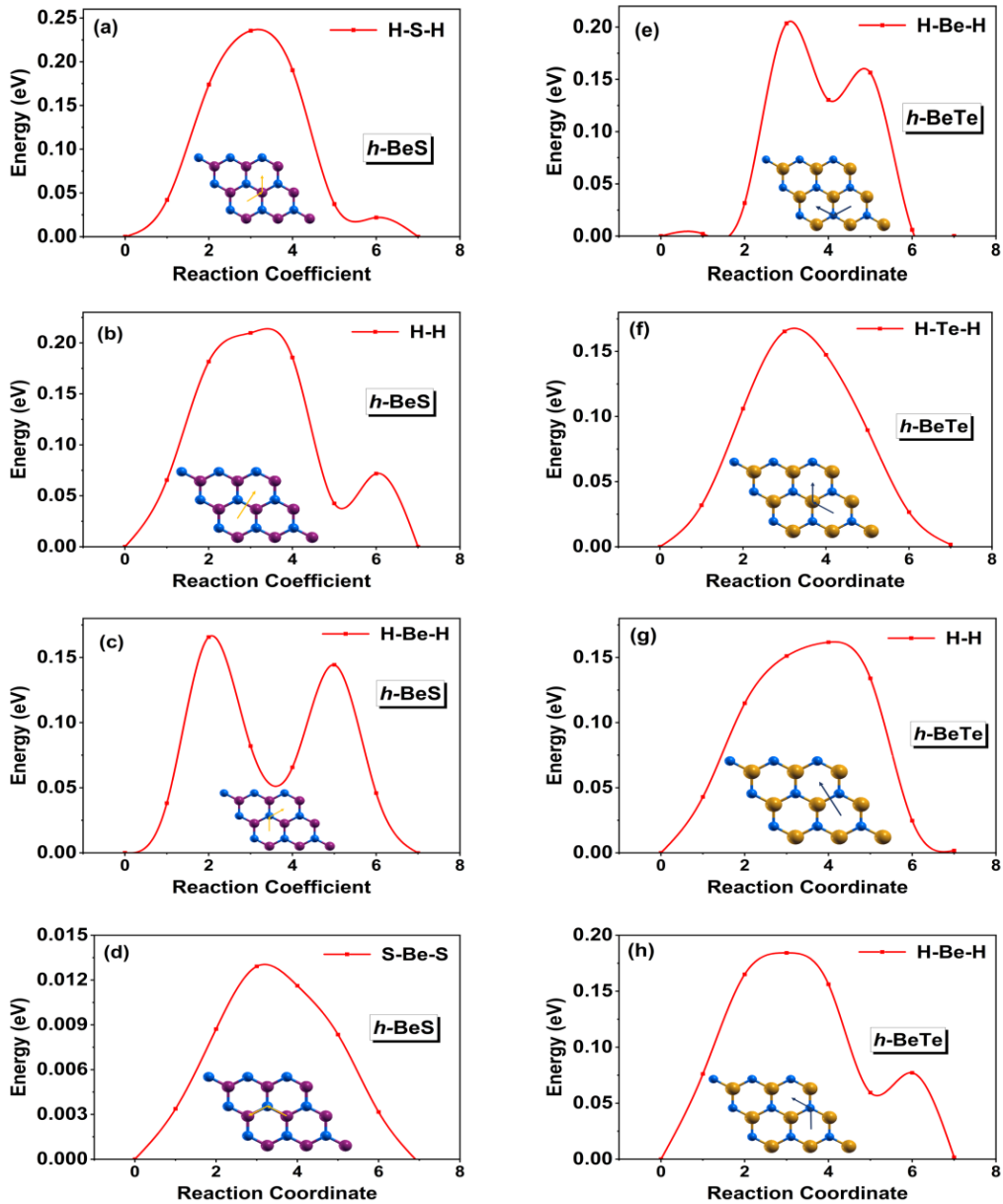

**Figure 6:** Diffusion barrier profile for *h*-BeS (a-c) Li atom, (d) Na atom and for *h*-BeTe (e) Li atom, (f-h) Na atom for specific pathways. Inset to respective figures, show the path for the migration of the alkali metal atom in the supercell structure.

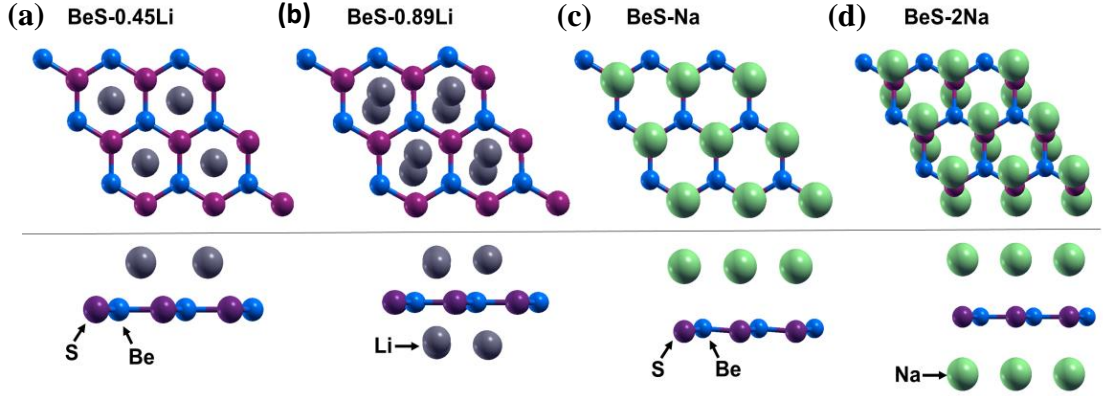

**Figure-7:** Top view of metal-adsorbed monolayer systems (a) BeS-0.45Li, (b) BeS-0.89Li, (c) BeS-Na, (d) BeS-2Na.

Average *OCV* was calculated using the half-cell reaction's Gibbs free energy ($\Delta G_f$) change. $\Delta G_f$ can be represented by $\Delta G_f = \Delta E_f + P\Delta V - T\Delta S$. Since the second term (*P$\Delta$V*) is negligible [48,49], and the last term, average entropy (*T$\Delta$S*) is approximately 25 meV [48,50] at room temperature, the $\Delta G_f$ can be roughly equated to the internal energy ($\Delta E_f$) [26,48]. We get *OCV* = - ($\Delta E_f$ / *xe*), which can be calculated using Eq. (2) [26,29].

$$OCV = [-E_{BeS/Te-M} - E_{BeS/Te} - xE_M] / xe \qquad (2)$$

where $E_{BeS/Te-M}$, $E_{BeS/Te}$, $E_M$, have been defined earlier in section 3.1, and $x$ represents the concentration of adsorbed atoms on the 3×3×1 supercell of BeS and BeTe monolayers, and *e* is the universal electronic charge. The calculated *OCV* values for BeS-0.45Li and BeS-0.89Li are 1.53 V and 1.88 V, respectively, whereas those of 1.03 V and 1.05 V are found for BeS-Na and BeS-2Na, respectively. These metal-adsorbed monolayer systems are shown in Fig. 7 (a-d). Similarly, for BeTe-0.45Li and BeTe-0.89Li, the calculated *OCV* values are 1.40 V and 2.12 V, respectively. Further, for the case of BeTe-0.45Na and BeTe-0.89Na, the *OCV* values are 0.95 V and 1.23 V, respectively. Understanding that theoretical and experimental viewpoints differ when classifying materials as anode and cathode is crucial. Theoretically, the *OCV* values between 0.1 – 1.25 eV generally indicate the anodic behaviour, while values over 1.25 eV indicate the cathodic behaviour [43,47,51,52]. Graphite, which has an experimental value of *OCV* 0.1 to 0.2 V [53], is the most commonly used anode material. The *OCV* value of another anode material, lithium titanium oxide, is 1.55 V [54], whereas cathode materials LiFePO$_4$ [55] and lithium manganese oxide [56] have an OCV in the range of 3.45 – 3.50 and 3.5 – 4.2 V, respectively. For the present study, the average value of *OCV* for BeS-Li and BeS-Na is 1.70 V and 1.04 V, respectively, whereas for BeTe-Li and BeTe-Na is 1.77 V and 1.09 V, respectively. Hence, from an experimental perspective, we can conclude that BeS and BeTe monolayers can serve as cathodes for LIBs and anodes for NIBs. Table 4 summarizes the comparison of this work with earlier reported findings.

**Table 4:** Comparison of adsorption energy, diffusion energy and *OCV* for these electrode materials for LIB and NIB.

| Material | Adsorption Energy (eV) | | Diffusion Energy (eV) | | *OCV* (V) | | Reference |
|---|---|---|---|---|---|---|---|
| | **Li** | **Na** | **Li** | **Na** | **Li** | **Na** | |
| *h*-BeS | -0.60 | -0.16 | 0.16 | 0.01 | 1.70 | 1.04 | This work |
| *h*-BeTe | -1.19 | -0.78 | 0.20 | 0.16 | 1.77 | 1.09 | This work |
| $Ti_2BN_2$ | -2.79 | -1.95 | 0.44 | 0.34 | 0.93 | 0.27 | [18] |
| $BSi_4$ | -1.52 | -1.38 | 0.33 | 0.22 | 0.80 | 0.52 | [57] |
| $Cr_2B_2F_2$ | -2.60 | -2.40 | 0.13 | 0.14 | 1.99 | 0.71 | [58] |
| $Cr_2B_2Cl_2$ | -1.55 | -1.25 | 0.20 | 0.20 | 0.98 | 0.29 | [58] |
| *h*-BAs | -0.42 | -0.32 | 0.52 | 0.25 | 0.49 | 0.35 | [59] |
| $Ca_2C$ | -2.24 | -2.84 | 0.03 | 0.06 | 0.10 | 0.24 | [60] |
| $P_3S$ | −5.52 | −5.62 | 0.74 | 0.26 | 1.75 | 1.84 | [47] |
| $C_3S$ | −5.02 | −4.90 | 0.68 | 0.30 | 1.06 | 0.50 | [47] |
| FeSe | -2.75 | -2.29 | 0.16 | 0.13 | 0.25 | 0.17 | [48] |
| $Mg_2C$ | -1.95 | -1.00 | 0.70 | 0.08 | 0.37 | 0.50 | [61] |

Additionally, the storage capacity of any cell, which is an essential parameter, can influence the viability of any cell. This parameter is also greatly affected by the choice of electrode material and the electrochemical reactions within the cell, and its maximum theoretical value can be determined using Eq. (3) [48,62],

$$C = \frac{x_{max} \times F}{M} \qquad (3)$$

where $x_{max}$ is the maximum number of Li/Na atoms adsorbed on the monolayer, *F* is the Faraday constant (26810 mAh/mol), and *M* is the molar mass of the respective pristine monolayer. For BeS-0.89Li and BeS-2Na, the computed value of *C* is 580 and 1305 mAh g$^{-1}$, respectively, whereas those for BeTe-0.89Li and BeTe-0.89Na systems are 174 mAh g$^{-1}$. These values are comparatively higher than previously reported values for other electrode materials such as graphite (372 mAh g$^{-1}$ for Li [63], 284 mAh g$^{-1}$ for Na [63]); $V_3C_2$ (606 mAh g$^{-1}$ for Na [64]); FeSe (473 mAh g$^{-1}$ for Na [48]); and $Mo_2C$ (526 mAh g$^{-1}$ for Li [22], for Na 132 mAh g$^{-1}$ [22]) and $MoS_2$ (146 mAh g$^{-1}$ for Na [65]).

## 4. Conclusion

In summary, the viability of stable *h*-BeS and *h*-BeTe monolayers as potential electrode choices for LIBs and NIBs has been theoretically examined for the first time. The negative formation energies, positive phonon frequencies and small fluctuations of total energy in AIMD simulations ensure these monolayers' structural, dynamical and thermodynamic stability. After the absorption of Li/Na atoms, monolayers change their nature from semiconductor to metallic, which is also confirmed by thermal transport investigations. A low diffusion barrier in the range 0.01 – 0.16 eV is calculated for the migration of alkali atoms along the preferred paths on these monolayers, ensuring a fast diffusion capability required for multiple charging-discharging cycles. Based on the estimated values of *OCV*, we infer that the BeS and BeTe monolayers serve as a cathode for LIBs and an anode for NIBs. The maximum value for the theoretical storage capacity is 580 and 1305 mAh g$^{-1}$ for Li and Na adsorbed BeS monolayer, respectively, whereas that of 174 mAh g$^{-1}$ for BeTe monolayer for Li or Na adsorption.

**Acknowledgement**

We acknowledge the National Supercomputing Mission (NSM) for providing access to the computing resources of 'PARAM RUDRA' located at Aruna Asaf Ali Marg, near Vasant Kunj, New Delhi – 110067, which is implemented by C-DAC and supported by the Ministry of Electronics and Information Technology (MeitY) and the Department of Science and Technology (DST), Government of India. We thank Prof. Debesh R. Roy and Aditya P. Patel, Department of Physics, SVNIT Surat, India, for providing access to VASP software and carrying out AIMD simulations and related discussions. HP gratefully acknowledges SVNIT Surat for the Institute Seed Research Grant (2021-22/DOP/04).

**Conflicts of Interest**

The authors declare no conflicts of interest.

**Data Availability Statement**

The data that support the findings of this study are available from the corresponding author upon reasonable request.

49. M.K. Aydinol, A.F. Kohan, G. Ceder, K. Cho, J. Joannopoulos, *Ab initio* study of lithium intercalation in metal oxides and metal dichalcogenides, Phys. Rev. B 56 (1997) 1354–1365. https://doi.org/10.1103/PhysRevB.56.1354.
50. J. Yang, J. Wang, X. Dong, L. Zhu, D. Hou, W. Zeng, J. Wang, The potential application of $VS_2$ as an electrode material for Mg ion battery: A DFT study, Applied Surface Science 544 (2021) 148775. https://doi.org/10.1016/j.apsusc.2020.148775.
51. B. Ball, P. Sarkar, Tuning the structural skeleton of a phenanthroline-based covalent organic framework for better electrochemical performance as a cathode material for Zn-ion batteries: a theoretical exploration, Phys. Chem. Chem. Phys. 23 (2021) 12644–12653. https://doi.org/10.1039/D1CP01209F.
52. H.R. Jiang, W. Shyy, M. Liu, L. Wei, M.C. Wu, T.S. Zhao, Boron phosphide monolayer as a potential anode material for alkali metal-based batteries, J. Mater. Chem. A 5 (2017) 672–679. https://doi.org/10.1039/C6TA09264K.
53. M. Kim, D.C. Robertson, D.W. Dees, K.P. Yao, W. Lu, S.E. Trask, J.T. Kirner, I. Bloom, Estimating the Diffusion Coefficient of Lithium in Graphite: Extremely Fast Charging and a Comparison of Data Analysis Techniques, J. Electrochem. Soc. 168 (2021) 070506. https://doi.org/10.1149/1945-7111/ac0d4f.
54. S. Yi, B. Wang, Z. Chen, R. Wang, D. Wang, A study on $LiFePO_4$/graphite cells with built-in $Li_4Ti_5O_{12}$ reference electrodes, RSC Adv. 8 (2018) 18597–18603. https://doi.org/10.1039/C8RA03062F.
55. S. Kanungo, A. Bhattacharjee, N. Bahadursha, A. Ghosh, Comparative Analysis of $LiMPO_4$ (M = Fe, Co, Cr, Mn, V) as Cathode Materials for Lithium-Ion Battery Applications—A First-Principle-Based Theoretical Approach, Nanomaterials 12 (2022) 3266. https://doi.org/10.3390/nano12193266.
56. N. Somakettarin, T. Funaki, Study on Factors for Accurate Open Circuit Voltage Characterizations in Mn-Type Li-Ion Batteries, Batteries 3 (2017) 8. https://doi.org/10.3390/batteries3010008.
57. J. Du, H. Lin, Y. Huang, Conductive $BSi_4$ monolayer with superior electrochemical performance for alkali metal ion batteries, Materials Science in Semiconductor Processing 172 (2024) 108086. https://doi.org/10.1016/j.mssp.2023.108086.
58. A.R. Nirjhar, S.J. Tan-Ema, M.A. Sahriar, Md.N.A. Dipon, Mohd.R. Hasan Abed, K.Md. Shorowordi, S. Ahmed, Tuning the electrochemical performance of $Cr_2B_2$ MBene anodes for Li and Na-ion batteries through F and Cl-functionalization: A DFT and AIMD study, Colloids and Surfaces A: Physicochemical and Engineering Aspects 684 (2024) 133194. https://doi.org/10.1016/j.colsurfa.2024.133194.
59. N. Khossossi, A. Banerjee, Y. Benhouria, I. Essaoudi, A. Ainane, R. Ahuja, Ab initio study of a 2D h-BAs monolayer: a promising anode material for alkali-metal ion batteries, Phys. Chem. Chem. Phys. 21 (2019) 18328–18337. https://doi.org/10.1039/C9CP03242H.
60. K. Rajput, V. Kumar, S. Thomas, M.A. Zaeem, D.R. Roy, $Ca_2C$ MXene monolayer as a superior anode for metal-ion batteries, 2D Mater. 8 (2021) 035015. https://doi.org/10.1088/2053-1583/abf233.
61. Y.Z. Chu, K.H. Yeoh, K.-H. Chew, A first-principles comparative study of lithium, sodium, potassium and calcium storage in two-dimensional $Mg_2C$, J. Phys.: Condens. Matter 33 (2021) 075002. https://doi.org/10.1088/1361-648X/abc807.
62. J M Tarascon, M Armand, Issues and challenges facing rechargeable lithium batteries, Nature 414 (2001) 359. https://doi.org/10.1038/35104644.
18